\documentclass[doublecol,figures]{epl2}

\usepackage{graphicx}
\usepackage{dcolumn}
\usepackage{bm}
\usepackage{hyperref}

\usepackage{color}
\usepackage{epsfig}
\usepackage{amsmath}
\usepackage{amsfonts}
\usepackage{amssymb}
\usepackage{units}
\usepackage{amssymb}
\usepackage{wasysym}

\title{Spin-squeezed atomic crystal}
\shorttitle{Spin-squeezed atomic crystal}
\author{Dariusz Kajtoch,$^{1,\, 2}$  Emilia Witkowska$^{2}$ and Alice Sinatra$^{1}$}
\shortauthor{D. Kajtoch, E. Witkowska, A. Sinatra}
\institute{$^{1}$Laboratoire Kastler Brossel, ENS-PSL, CNRS, UPMC-Sorbonne Universit\'e 
and Coll\`ege de France, Paris, France
$^{2}$Institute of Physics, PAS, Aleja Lotnikow 32/46, PL-02668 Warsaw, Poland}

\pacs{03.75.Gg}{Entanglement and decoherence in Bose-Einstein condensates}
\pacs{03.75.Kk}{ Bose-Einstein condensation dynamic properties} 
\pacs{42.50.Dv}{Squeezed states}
\pacs{06.30.Ft}{Time and frequency}

\abstract{
We propose a method to obtain a regular arrangement of two-level atoms in a three-dimensional optical lattice with unit filling, where all the atoms share internal state coherence and metrologically useful quantum correlations. Such a spin-squeezed atomic crystal is obtained by adiabatically raising an optical lattice in an interacting two-component Bose-Einstein condensate. The scheme could be directly implemented on a microwave transition with state-of-the art techniques and used in optical-lattice atomic clocks with bosonic atoms to strongly suppress the collisional shift and benefit from the spins quantum correlations at the same time.
}

\begin{document}
\date{9 February 2018}

\maketitle
\section{Introduction} 
One of the most successful applications of cold atoms is atomic clocks that provide the best time standards. 
The intrinsic uncertainty in the measured clock transition frequency, so called standard quantum limit $\Delta \omega \propto ({\cal T}\sqrt{N})^{-1}$ where ${\cal T}$ is the interrogation time and $N$ the number of atoms, has been already reached in atomic fountains using a microwave transition \cite{PhysRevLett.82.4619}.
{Increasing the atom number $N$ and the interrogation time ${\cal T}$ in order to lower the value of the standard quantum limit is not straightforward in an atomic fountain. Indeed too large densities introduce
an atom-number dependent frequency shift due to atomic interactions (collisional shift) and the interrogation time is limited by the fact that the cold cloud is in free expansion. Optical lattices that allow to confine and interrogate simultaneously a large number of atoms for longer times offer in this respect a crucial advantage \cite{nature12941,PhysRevA.81.023402,nphys1108,1882-0786-10-7-072801,PhysRevLett.101.220801,PhysRevLett.106.063002} that, combined with the use of a larger reference frequency (optical transition) already outperforms the 
regular frequency standards \cite{Ludlow1805,Hinkley1215,nature12941}. Interestingly, many-body physics of cold atoms in an optical lattice opens new perspectives that can further improve atomic clocks.
In particular, the configuration with one atom per site of an optical lattice is a mean to suppress the collisional shift, maximizing at the same time the density, both
for bosonic \cite{PhysRevA.81.023402,nphys1108,1882-0786-10-7-072801} and also for fermionic atomic clocks \cite{JYe2017}.
On the top of these rapid developments, in this paper we envisage to introduce
well designed quantum correlations among the atoms internal states, known as {\it spin-squeezing} \cite{PhysRevA.50.67,nature08988,nature08919,PhysRevLett.104.073602,Hosten2016,PhysRevLett.116.093602}, in order to beat the standard quantum limit and push even further the extraordinary precision of optical-lattice clocks  in the long term. To this aim
we generalize a spin-squeezing scheme successfully implemented in bimodal Bose-Einstein condensates \cite{nature08988,nature08919} to the spatially multimode case of an optical lattice.

Our idea is to raise the optical lattice in an interacting bimodal Bose-Einstein condensate, adiabatically bringing the system from the superfluid to the Mott-insulator phase \cite{Greiner2002,PhysRevA.79.041604,PhysRevA.77.011603}. The correlations among the atomic spins build up in the superfluid phase where ``each atom sees each atom" similarly to what happens in the absence of lattice where all the atoms share the same spatial mode. As the system approaches the Mott transition, the squeezing dynamics slows down to finally stop completely. While the condensate is destroyed as the atoms get localized in the lattice sites, the spin-squeezing survives and it is stored in the Mott-insulator phase. To the advantage of having a close packed ensemble atoms, where atomic interactions are highly suppressed
meeting the basic requirements for an atomic clock, we then add the advantage of quantum correlations allowing for metrological gain with respect to independent atoms.
We analyze in detail the scheme for a microwave transition relevant to microwave 
trapped-atoms bosonic clocks \cite{PhysRevLett.101.220801,ChinLett2009,PhysRevA.81.031611,PhysRevLett.106.063002,PhysRevLett.117.150801}, for which a proof-of-principle experiment may be preformed with state of the art techniques. }

Besides atomic clocks, the scheme can be used to ``extract" the entanglement among the atoms that are initially in a common spatial mode \cite{PhysRevLett.112.150501}, to obtain a state where each entangled atom can be individually addressed and manipulated with a quantum gas microscope \cite{Bakr2009,Sherson2010}. Our ``spin-squeezed crystal" is then a platform 
that allows a complete characterization of the entangled state \cite{PhysRevA.67.052315} and could be used as a resource 
in the emerging field of quantum networks and multiparameter estimation.

\begin{figure}[]
\includegraphics[width=\linewidth]{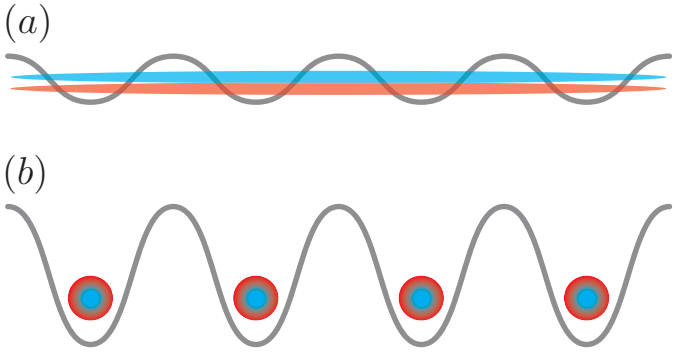}
\caption{
(a) Initially, a Bose-Einstein condensate of atoms in an internal state $a$ is prepared in a shallow 3D optical lattice. 
At time $t=0$ an electromagnetic $\pi/2$-pulse puts each atom in a coherent superposition two internal states $a$ and $b$, and the binary atomic interactions between cold atoms start the generation of squeezing in the system \cite{PhysRevA.47.5138,Sorensen2001}. 
(b) Simultaneously, the lattice height is gradually increased, in such a way that the system enters the Mott-insulator phase at the ``best squeezing time" $t_{\rm best}$ for which squeezing is the largest, thus freezing the ``best squeezing" $\xi^{2}_{\rm best} = \xi^2(t_{\rm best})$ (see equation (\ref{eq:spin_squeezing})) in the Mott-insulator phase. }
\label{fig:lattice}
\end{figure}

\section{The Model} 
The protocol is sketched in Fig.~\ref{fig:lattice}. 
We consider a two component condensate with repulsive interactions, 
with symmetric coupling constants describing $s$-wave interactions between atoms in the two states $g_{aa} = g_{bb}$, and an adjustable interspecies coupling \cite{PhysRevLett.82.1975,nature08919}
{in the phase-mixed regime $g_{ab}<g_{aa}$ \cite{PhysRevLett.81.5718}}; $g_{ij}=4\pi \hbar^2 a_{ij}/m$ where $a_{ij}$ is the $s$-wave scattering length for one atom in state $i$ and one in state $j$, $m$ is the atomic mass and $\hbar$ is the Planck constant.
The system Hamiltonian is 
\begin{multline}
\hat{H} = \sum_{\sigma=a,b} \left\{ \int d^3r \: \hat{\Psi}_\sigma^\dagger({\bf r}) \left[ -\frac{\hbar^2}{2m} \nabla^2 + V({\bf r}) \right] \hat{\Psi}_\sigma({\bf r})  \right.\\ \left.
+  \frac{g_{\sigma \sigma}}{2} \int d^3r \: \hat{\Psi}_\sigma^\dagger({\bf r})\hat{\Psi}_\sigma^\dagger({\bf r})\hat{\Psi}_\sigma({\bf r})\hat{\Psi}_\sigma({\bf r}) \right\} \\
+  g_{ab} \int d^3r \: \hat{\Psi}_a^\dagger({\bf r})\hat{\Psi}_b^\dagger({\bf r})\hat{\Psi}_b({\bf r})\hat{\Psi}_a({\bf r}) ,
\label{eq:Hamiltonian}
\end{multline}
where $\hat{\Psi}_a$ and $\hat{\Psi}_b$ are the bosonic field operators for atoms in the state $a$ and $b$ respectively satisfying usual commutation relations,
\begin{equation}
[\hat{\Psi}_\sigma({\bf r}),\hat{\Psi}_{\sigma'}^\dagger({\bf r'})]=\delta({\bf r}-{\bf r'})\delta_{\sigma \sigma'} .
\end{equation}
 The system is confined in a 3D uniform optical lattice described by the periodic potential
\begin{equation}
V({\bf r})=V_0 \sum\limits_{\alpha = x,y,z} \sin^{2}(k \alpha)
\label{eq:pot_per}
\end{equation}
where $k = 2\pi/\lambda$ is the lattice wavenumber and $\lambda/2$ is the lattice period, with the number of lattice sites $M$ equal to the number of atoms $N$. {Uniform systems can be realized in a flat bottom potential that is now possible to produce in the laboratory \cite{PhysRevLett.110.200406,Zwirlein}. The fact that the potential walls are not infinitely steep} releases the constraint of having exactly a filling factor of one to reach the Mott transition \cite{Lundh2008}. After each atom is prepared in a coherent superposition of internal states $a$ and $b$ by an electromagnetic 
$\pi/2$ pulse, the lattice depth is linearly increased in time
\begin{equation}
V_0(t) = V_{\rm init} + (V_{c} - V_{\rm init})t/t_{\rm best}
\label{eq:ramp}
\end{equation}
from an initial value $V_{\rm init}$ to the critical value $V_c$ to reach the Mott transition, on a time scale given by the best squeezing time $t_{\rm best}$.
We assume that the system dynamics is confined to the lowest Bloch band and it is described by the two-component Bose-Hubbard model \cite{PhysRevA.81.043620} with time-dependent hopping $J(V_0)$ and interaction terms $U_\sigma(V_0)$, $U_{ab}(V_0)$.
We quantify the spin-squeezing using the parameter \cite{PhysRevA.46.R6797, PhysRevA.50.67}
\begin{equation}\label{eq:spin_squeezing}
\xi^2 = \frac{N \langle \Delta \hat{S}^2_{\perp}\rangle_{\rm min}}{\langle S\rangle^2},
\end{equation}
where $\langle S\rangle$ is the length of the mean collective spin and $\langle \Delta \hat{S}^2_{\perp}\rangle_{\rm min}$ is the minimal variance of the spin orthogonally to the mean spin direction. 
The collective spin operators, $\hat{S}_\pm  = \hat{S}_x \pm i \hat{S}_y = \int d^3r \hat{\Psi}_a^{\dagger}(\mathbf{r})\hat{\Psi}_{b}(\mathbf{r})$, $ \hat{S}_z = (\hat{N}_a - \hat{N}_b)/2$ where $\hat{N}_\sigma$ is the atom number operator in the internal state $\sigma$, give access to spin-squeezing as a function of time.

\section{Initial state and evolution} 
Starting from a condensate  in internal state $a$, a mixing $\pi/2$ pulse puts the system in a superposition of Fock states with $N_a$ atoms in the internal state $a$ and $N_b=N-N_a$ atoms in $b$
$|\Psi^0_{N_a, N_b}(0^+): N_a, N_b \rangle$ where we assume that each Fock state is in the spatial  ground state for the corresponding atom number in the shallow lattice. We thus neglect the thermal excitations initially present in the system and the excitations created by the mixing pulse. A clean way to avoid any excitation in the experiment would be to perform the pulse for an ideal gas at zero temperature in a completely flat potential, and subsequently ramp up adiabatically the interactions \cite{PhysRevA.95.063609} and the lattice. The very stringent condition of having no excitations at all is however not necessary in practice for our proposal. Based on our studies of spin-squeezing at finite temperature \cite{PhysRevLett.107.060404,Sinatra2012}, we expect that the final squeezing will not be affected as long as the initial non-condensed fraction before the pulse is much smaller than the targeted squeezing, which can be achieved in the weakly interacting limit at sufficiently low temperature.

In the adiabatic approximation, during the evolution following the mixing pulse, each Fock state of the prepared superposition remains in an instantaneous ground state and picks up a time-dependent phase factor determined by the ground state energy $E_0(N_a, N_b, t)$
\begin{multline}
 | \psi(t) \rangle = 2^{-N/2} \sum\limits_{N_a = 0}^{N}  \sqrt{\binom{N}{N_a}} e^{-\frac{i}{\hbar} \int\limits_{0}^{t} dt' E_0(N_a, N-N_a, t') } \\
 \times |\Psi^0_{N_a, N-N_a}(t): N_a, N-N_a \rangle.
 \label{eq:time_evolution}
\end{multline} 
{Equation (\ref{eq:time_evolution}) for the system evolution is exact in the adiabatic approximation, as long as 
$\Psi^0_{N_a, N_b}$ and $E_0(N_a, N_b)$ refer to the exact ground-state of the system with $N_a$ atoms in the internal state $a$ and $N_b$ atoms in state $b$.
The $N$-body ground state for two-component bosons on a lattice exhibits non-trivial correlations \cite{SvistunovCF,Altman2CBH} that entangle the internal and external degrees of freedom in equation (\ref{eq:time_evolution}) which would render the calculation of the squeezing parameter a formidable task.  We can however simplify the problem by introducing two approximations.
(i)} We assume that the squeezing is measured after raising the lattice to reach the deep Mott phase with one atom per site in the limit of vanishing tunnel coupling, which allows to disentangle the internal and external degrees of freedom in state (\ref{eq:time_evolution}). A consequence of this fact is that the action of the collective spin operators on the Fock states then obeys simple rules, e.g. 
$\hat{S}_+|\Psi^0_{N_a, N_b}: N_a, N_b \rangle \approx \sqrt{(N_a+1)N_b} |\Psi^0_{N_a+1, N_b-1} : N_a+1, N_b-1 \rangle$, as for the two-mode Fock state where only internal degrees of freedom are considered.
{(ii) We evaluate the energies $E_0(N_a, N_b)$ entering in the phase factors in (\ref{eq:time_evolution}) in the mean field approximation for the external degrees of freedom of the atoms,}
using the Gutzwiller method \cite{PhysRevB.44.10328,PhysRevA.71.043601,PhysRevLett.89.040402,PhysRevA.81.043620}. Precisely, the total mean-field energy is minimized numerically in the subspace of fixed mean number of atoms in each component, i.e. $\langle \hat{N}_a \rangle = N_a$ and $\langle \hat{N}_b \rangle = N_b$, by a projection method similarly to Ref.\cite{PhysRevE.78.066704}. 
{This approximation is accurate enough in 3D and captures the transition to the Mott state marked by the suppression of on-site atom number fluctuations for integer filling and a threshold value of the $U/J$ parameter of the Bose-Hubbard model.}
The results for the on-site correlation functions $g^{(2)}_{aa} = \langle \hat{a}^{\dagger}_i \hat{a}^{\dagger}_i
\hat{a}_i \hat{a}_i \rangle / \langle \hat{a}^{\dagger}_i \hat{a}_i
\rangle^2$ and $g^{(2)}_{ab} = \langle \hat{a}^{\dagger}_i \hat{b}^{\dagger}_i
\hat{a}_i \hat{b}_i \rangle / \langle \hat{a}^{\dagger}_i \hat{a}_i
\rangle \langle \hat{b}^{\dagger}_i \hat{b}_i \rangle$ and for the squeezing parameter $\xi^2$ as a function of time are shown in Fig.~\ref{fig:fig1}(a) and in Fig.~\ref{fig:fig1}(b) (green solid line) respectively. 
{The time where the on-site $g^{(2)}$ functions go to zero, indicated with a vertical dotted line, marks the transition to the Mott phase. At this point the atoms cease to ``feel each other" via the mean field interaction. As a consequence the squeezing dynamics stops and the spin correlations that built up in the superfluid phase are ``frozen" in the Mott phase.}

\section{{Effective $\chi \hat{S}_z^2$ model in the superfluid phase}}
If the atom number is large, one can expand the ground state energy $E_0(N_a, N_b, t)$ in the phase factor of each Fock state in~\eqref{eq:time_evolution} around $\langle \hat{N}_a \rangle= \langle \hat{N}_b \rangle = N/2$ up to the second order~\cite{Sinatra1998}
\begin{equation}\label{eq:sz_square}
\frac{1}{\hbar} \int\limits_{0}^{t} dt' E_0(N_a, N_b, t') \simeq \Phi_0(N,t) + T(t) \frac{(N_a - N_b)^2}{4} \,.
\end{equation}
There is no linear term in (\ref{eq:sz_square}) because of the $a-b$ symmetric situation we consider, and the phase factor $\Phi_0(N,t)$ depending on the total number of atoms does not play any role in the spin dynamics and can be neglected. 
Introducing the chemical potentials $\mu_{\sigma}(t) = \partial_{N_{\sigma}}E_0(N_a, N_b, t)$ and  the parameter $\chi(t) = \frac{1}{2} (\partial_{N_a}-\partial_{N_b})[\mu_a(t) - \mu_b(t)]/\hbar$, the function $T(t)$ in equation (\ref{eq:sz_square}) has the form
\begin{equation}\label{eq:T(t)}
T(t) = \int\limits_{0}^{t}dt' \chi(t') \,.
\end{equation}
By changing the time variable from $t$ to the dimensionless $T(t)$, one then recovers the one-axis twisting  (OAT) model \cite{PhysRevA.47.5138} with an Hamiltonian proportional to $\hat{S}_z^2$.
In particular, in the large $N$ limit, the squeezing optimized over time $\xi^{2}_{\rm best} = \xi^2(t_{\rm best})$ is $\xi^{2}_{\rm best}\simeq \revision{ \frac{3^{2/3}}{2} } \frac{1}{N^{2/3}}$ and the best squeezing time is
\begin{equation}
t_{\rm best} \simeq \frac{3^{1/6}}{N^{2/3}} \left[ \frac{1}{V_c - V_{\rm init}} \int\limits_{V_{\rm init}}^{V_c} dV_0\ \chi(V_0) \right]^{-1}
\label{eq:tb_scaling}
\end{equation}
where we used $ T(t_{\rm best}) \simeq 3^{1/6}/N^{2/3}$ and the linear ramp (\ref{eq:ramp}).
We introduce the term ``dynamic-OAT model" for the quadratic approximation (\ref{eq:sz_square}) in the phase factors, as opposite to ``static-OAT model" for which we directly take a time independent Hamiltonian $\hat{\mathcal{H}} = \hbar \chi(0) \hat{S}_z^2$ corresponding to the initial conditions after the pulse. 
\begin{figure}[h!]
 \centering
 \includegraphics[width=\linewidth]{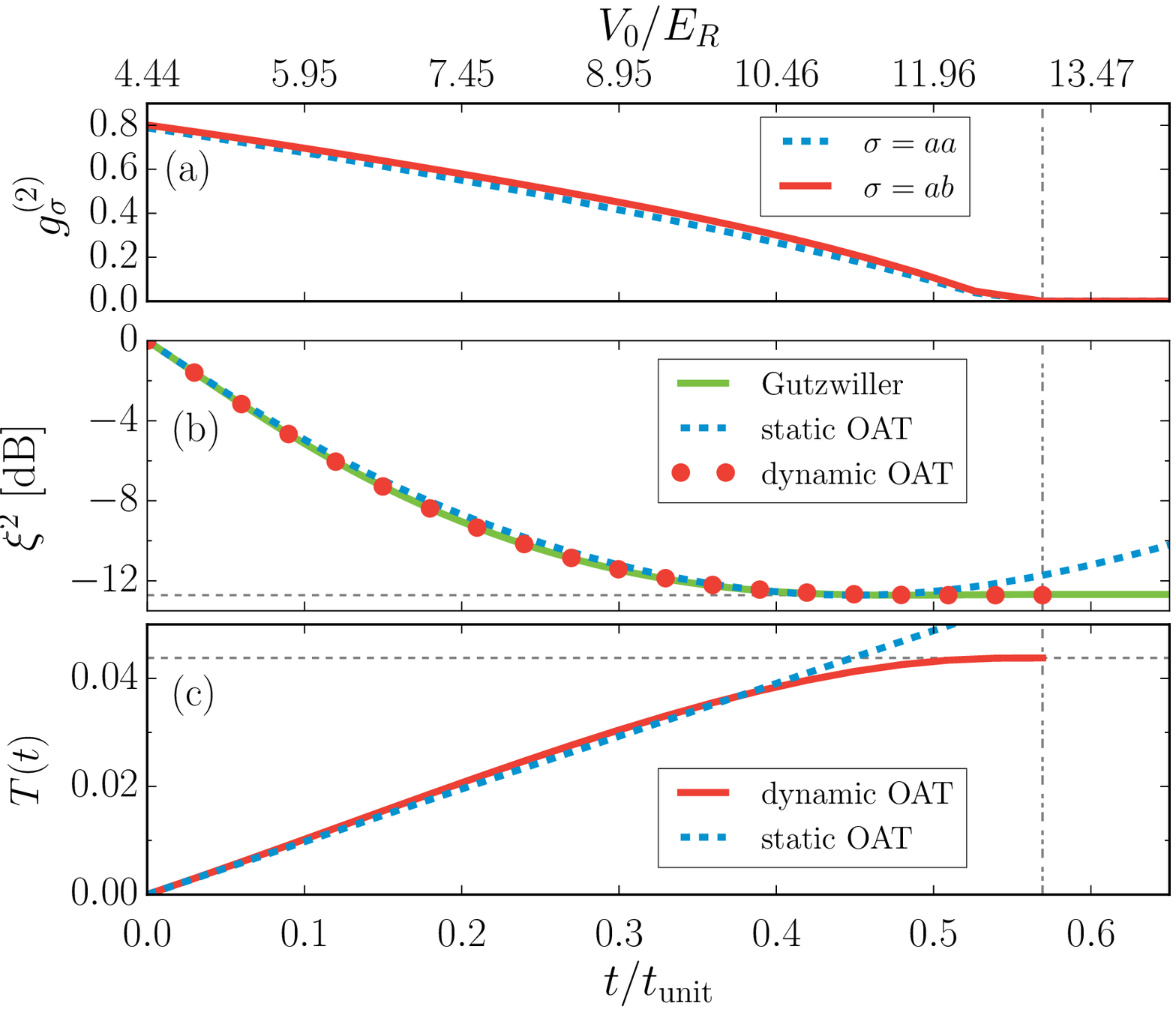}
 \includegraphics[width=\linewidth]{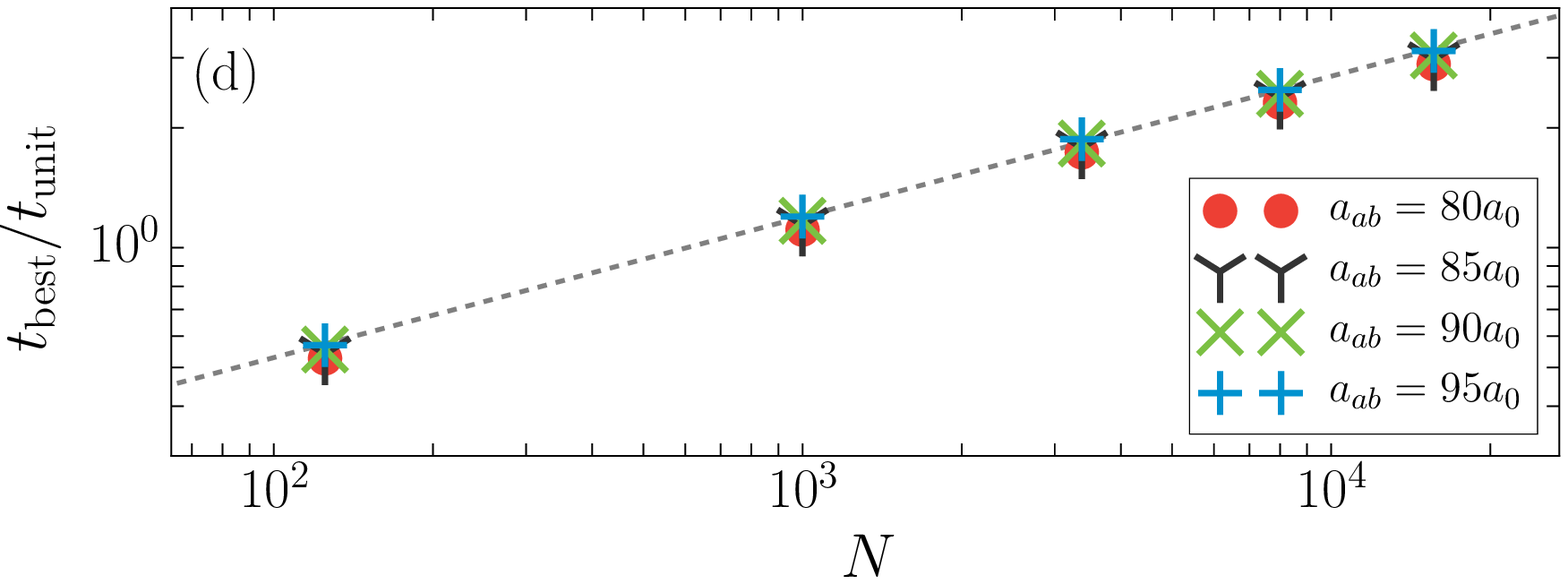}
 \caption{\label{fig:fig1} 
 (Color online) Numerical results:
(a) On-site two-body correlation functions 
$g^{(2)}_{aa} = \langle \hat{a}^{\dagger}_i \hat{a}^{\dagger}_i
\hat{a}_i \hat{a}_i \rangle / \langle \hat{a}^{\dagger}_i \hat{a}_i
\rangle^2$ and $g^{(2)}_{ab} = \langle \hat{a}^{\dagger}_i \hat{b}^{\dagger}_i
\hat{a}_i \hat{b}_i \rangle / \langle \hat{a}^{\dagger}_i \hat{a}_i
\rangle \langle \hat{b}^{\dagger}_i \hat{b}_i \rangle$
calculated for the central Fock state $|N_a =N/2, N_b = N/2\rangle$. 
(b) Spin-squeezing parameter $\xi^2$ as a function of $t/t_{\rm unit}$ (bottom x-axis) and $V_0/E_R$ (top x-axis) for the linear ramp; $E_R=2\pi^2\hbar^2/(m\lambda^2)$ is the recoil energy and $\lambda/2$ the lattice period. $t_{\rm unit}^{-1}=(a_{aa} + a_{bb} - 2 a_{ab})E_R/\hbar \lambda$. The green solid line shows numerical results with $E_0$ 
in ~\eqref{eq:time_evolution} calculated with the Gutzwiller method, while the red dots result from the dynamic-OAT model (\ref{eq:sz_square}). 
Predictions of the static-OAT model (see text) are shown for comparison (blue dashed lines). 
(c) $T(t)$ for the dynamic-OAT model (red solid line) and the static-OAT model (blue dashed line).
Parameters are $N = 125$, $a_{aa} = a_{bb} = 100.4 a_0$ and $a_{ab} = 95 a_0$, where $a_0$ is the Bohr radius. The dashed horizontal lines in (b) and in (c) represent respectively $\xi^{2}_{\rm best}$ and $T(t_{\rm best})$ of the OAT model for $N = 125$.  By definition, the dynamic curves touch the dashed horizontal lines at the best squeezing time $t_{\rm best}$. 
(d)  Scaling of the best squeezing time $t_{\rm best}$ with $N$ for $a_{aa} = a_{bb} = 100.4 a_0$ and different values of $a_{ab}$. The dashed line is a fit $t_{\rm best} \propto N^{\alpha}$ with $\alpha =0.353 \pm~0.004$.
}
\end{figure}
{Contrarily to the approach described in the previous section based on the numerical solution of equation~\eqref{eq:time_evolution}, the dynamic-OAT model is valid only in the superfluid phase where the derivatives entering the definition of $\chi(t)$ are well defined. It allows however to get a physical insight and simple analytical results. }
In Fig.~\ref{fig:fig1}(b), giving the time evolution of the squeezing parameter, we show that the dynamic-OAT model (red circles) is in agreement with the numerical solution of equation~\eqref{eq:time_evolution} (green solid line) up to the phase transition already for $N=125$. 
In Fig.~\ref{fig:fig1}(c) we show the corresponding time dependence of $T(t)$. {When approaching the transition, $\chi(t)$ tends to zero and $T(t)$, that is the effective time of the effective $\chi \hat{S}_z^2$ model,  tends to a constant. For optimal results, the Mott transition should occur at the best squeezing time, whose expression for large $N$ is given in equation (\ref{eq:tb_scaling}).} As expected, the squeezing dynamics is slower in the dynamic-OAT model than in static one.
We show the scaling of the best squeezing time with the total number of atoms $N$ in Fig.~\ref{fig:fig1}(d). A fit gives
$t_{\rm best} \propto N^{0.35}$, a slightly less favorable scaling than for the homogeneous static-OAT model, for which $\chi(0)^{\rm homo} = 8(a_{aa} + a_{bb} - 2 a_{ab})E_R/\pi \lambda \hbar N$ and $t_{\rm best} \propto N^{1/3}$. The same figure confirms that the dependence of $t_{\rm best}$ on the scattering lengths suggested by the homogeneous static-OAT model, approximately holds.

\section{\revision{Adiabaticity and beyond mean field effects}}

\revision{In our treatment based on equation (\ref{eq:time_evolution}), while we treat exactly the spin degrees of freedom that are in a quantum superposition in the initial state, we perform two main approximations concerning the external degrees of freedom of the atoms. First, we assume adiabaticity as the optical lattice lattice is raised, second we use the Gutwiller approximation to evaluate the ground state energy for a given spin Fock state.  In this paragraph, we extend our analysis beyond mean field. 
As the squeezing  in our scheme develops essentially in the superfluid phase, we first concentrate on this phase and derive the validity conditions of the adiabatic approximation at heart of our treatment using the number-conserving Bogoliubov theory~\cite{PhysRevA.57.3008,PhysRevA.56.1414}. Then, in the end of the pararagraph, we discuss the effect of residual density fluctuations in the Mott state.}

In the superfluid phase, where delocalization of atoms is energetically more favorable, a large majority of particles occupy the zero quasi-momentum Bloch state of the lowest band giving rise to condensation in momentum space. When the depletion of the condensate is small, the Bogoliubov method can capture the physical properties of the Bose-Hubbard model \cite{PhysRevA.63.053601}, although it cannot be pushed too far towards the phase transition boundary \cite{PhysRevA.84.053613, PhysRevA.63.053601}. The starting point is the two-component Bose-Hubbard Hamiltonian \cite{PhysRevA.81.043620} written in quasi-momentum representation. This Hamiltonian has the same form as for a two-component Bose-Einstein condensate in free space\cite{1751-8121-41-14-145005} with the field operators $\hat{\Psi}_{\sigma}(\mathbf{r})$ expanded in a plane wave basis, provided the kinetic energy $\epsilon_{\mathbf{q}} = \hbar^2 \mathbf{q}^2/2 m$ and the coupling constants $g_\sigma$, $g_{ab}$ are replaced by $\epsilon_{\mathbf{q}}(t) = -2J(V_0) \sum_{\gamma} \cos(\mathbf{q} \cdot \mathbf{e}_{\gamma})$ and $U_\sigma(V_0)$, $U_{ab}(V_0)$ respectively,  where $\mathbf{e}_{\gamma}$ are primitive lattice vectors. One then
introduces the number-conserving operators, $\hat{\Lambda}_{\mathbf{q},a} = \hat{a}_{\mathbf{0}}^{\dagger}\hat{a}_{\mathbf{q}}/\sqrt{N_a}$ and $\hat{\Lambda}_{\mathbf{q},b} = \hat{b}_{\mathbf{0}}^{\dagger}\hat{b}_{\mathbf{q}}/\sqrt{N_b}$ representing the non-condensed fields, and quadratizes the Hamiltonian in these fields. 
The quadratic Hamiltonian is diagonalized using the generalized Bogoliubov transformation 
\begin{align}
 \hat{\Lambda}_{\mathbf{q},\sigma}(t) = & u_{\mathbf{q},+}^{\sigma}(t) \hat{\beta}_{\mathbf{q},+}(t) + v_{\mathbf{q},+}^{\sigma}(t) \hat{\beta}^{\dagger}_{-\mathbf{q},+}(t) + \nonumber\\
 & u_{\mathbf{q},-}^{\sigma} \hat{\beta}_{\mathbf{q},-}(t) + v_{\mathbf{q},-}^{\sigma}(t) \hat{\beta}^{\dagger}_{-\mathbf{q},-}(t),
 \label{eq:bogoliubov_trans}
\end{align}
where $\hat{\beta}_{\mathbf{q},\pm}(t)$ are Bogoliubov quasi-particle operators satisfying bosonic commutation relations and $u_{\mathbf{q},\pm}^{\sigma}(t),\, v_{\mathbf{q},\pm}^{\sigma}(t)$ are known functions \cite{PhysRevLett.81.5718, 1751-8121-41-14-145005}. The Bogoliubov transformation in Eq.~\eqref{eq:bogoliubov_trans} is the same as in \cite{1751-8121-41-14-145005} with the same notations, provided that equations are expressed in terms of $\Delta E_{\mathbf{q}}(t) = \epsilon_{\mathbf{q}}(t) - \epsilon_{\mathbf{0}}(t)$. 

In order to calculate the number of excitations created by the lattice rump starting at zero temperature with no excitation, we write the Heisenberg equation of motion for $\hat{\beta}_{\mathbf{q}, \pm}(t)$, and proceed similarly to appendix C of Ref.\cite{PhysRevA.95.063609}. One can show that the Bogoliubov modes $\hat{\beta}_{\mathbf{q},+}(t)$ and 
$\hat{\beta}_{\mathbf{q},-}(t)$ evolve independently, and are coupled to modes $\hat{\beta}_{\mathbf{-q},+}(t)$ and 
$\hat{\beta}_{\mathbf{-q},-}(t)$ respectively
\begin{align}
i\hbar \frac{d}{dt}
\left(
 \begin{array}{c}
  \hat{\beta}_{\mathbf{q}, \pm}(t) \\
  \hat{\beta}^{\dagger}_{-\mathbf{q}, \pm}(t)
 \end{array}
\right)
 = &
\left(
 \begin{array}{cc}
  \hbar \omega_{\mathbf{q},\pm}(t) & -i\hbar \Omega_{\mathbf{q},\pm}(t) \\
  -i\hbar \Omega_{\mathbf{q},\pm}(t) & -\hbar \omega_{\mathbf{q},\pm}(t)
 \end{array}
\right) \times 
\nonumber \\[3mm]
& \left(
 \begin{array}{c}
  \hat{\beta}_{\mathbf{q}, \pm}(t) \\
  \hat{\beta}^{\dagger}_{-\mathbf{q}, \pm}(t)
 \end{array}
\right),
\end{align}
where the Bogoliubov spectrum is defined by
$(\hbar \omega_{\mathbf{q},\pm}/\Delta E_{\mathbf{q}})^2=1+\tilde{u}_a+\tilde{u}_b\pm \sqrt{(\tilde{u}_a - \tilde{u}_b)^2 + 4 \tilde{u}_{ab}}$ with $\tilde{u}_\sigma=U_\sigma n_\sigma/\Delta E_{\mathbf{q}}$ and $n_\sigma=N_\sigma/M$ for $\sigma=a,b$ \,, $\tilde{u}_{ab}=U_{ab}\sqrt{n_an_b}/\Delta E_{\mathbf{q}}$
and $\Omega_{\mathbf{q},\pm }(t) = \frac{1}{2}\frac{d}{dt}\log \left(
\frac{\Delta E_{\mathbf{q}}(t)}{\hbar \omega_{\mathbf{q},\pm}(t)}\right)$.
The total fraction of excitations
$\frac{1}{N}\sum_{\mathbf{q} \neq \mathbf{0}} n_{\mathbf{q}}^{\rm ex}(t)$ with $n_{\mathbf{q}}^{\rm ex}(t)=\sum_{\epsilon=\pm} \langle \Psi_{\rm bog}(0)| \hat{\beta}^{\dagger}_{\mathbf{q},\epsilon}(t)\hat{\beta}_{\mathbf{q},\epsilon}(t) 
 | \Psi_{\rm bog}(0) \rangle$ stays small, as long as the {(sufficient)} adiabaticity condition 
\begin{equation}\label{eq:two_adiabatic_cond}
|\Omega_{\mathbf{q},\pm}(t)| \ll |2 \omega_{\mathbf{q},\pm}(t)|
\end{equation}
is verified for each quasi-momentum $\mathbf{q}\neq \mathbf{0}$.
From the expression of $\Omega_{\mathbf{q},\pm}(t)$ and for the linear ramp (\ref{eq:ramp}) one gets the expression of the adiabatic time for each mode
\begin{equation}
t_{\rm adiab,\pm}^\mathbf{q} = \frac{(V_c - V_{\rm init}) \hbar E_R}{4 \Delta E_{\mathbf{q}}(V_0)} \left| \frac{d}{dV_0}\left( \frac{\Delta E_{\mathbf{q}}(V_0)}{\hbar \omega_{\mathbf{q},\pm}(V_0)}\right) \right| 
\end{equation}
which should be evaluated at $\mathbf{q} = (2\pi/N^{1/3} l, 0, 0)$ with the lattice spacing $l=\lambda/2$, where the condition~\eqref{eq:two_adiabatic_cond} is most stringent, and maximized over the ramp duration $V_0 \in [V_{\rm init}, V_c]$. In the large atom number limit one gets
\begin{equation}\label{eq:time_asymptotic}
t_{\rm adiab,\pm} \underset{N \rightarrow \infty}{\simeq} \frac{N^{1/3}}{2\pi} \frac{(V_c - V_{\rm init}) l}{4 J(V_0)} \left| \frac{d}{dV_0} \left( \frac{J(V_0)}{c_{\pm}(V_0)}\right) \right|,
\end{equation}
where $c_{\pm}^2=(l/\hbar)^2 J(V_0)((\hbar \omega_{\mathbf{q},\pm}/\Delta E_{\mathbf{q}})^2 - 1)\Delta E_{\mathbf{q}}$ is the sound velocity of the phonon-like Bogoliubov excitation branches. We see from Eq.~\eqref{eq:time_asymptotic} that the adiabatic time shares almost the same scaling with $N$ as the best squeezing time. Furthermore, the negative branch $t_{\rm adiab,-}$ is always larger than $t_{\rm adiab,+}$ as long as one of the components is not completely depleted. This implies that $t_{\rm adiab, -}$ alone sets the adiabatic time scale. We checked numerically that the condition $t_{\rm best} \gg t_{\rm adiab, -}$ holds for all parameters that we consider. For example, for $N=10^4$ and parameters in 
Fig.~\ref{fig:fig1}, $t_{\rm adiab}/t_{\rm best} \approx 0.032$.

\section{Decoherence}
\begin{figure}[]
\medskip
\includegraphics[width=\linewidth]{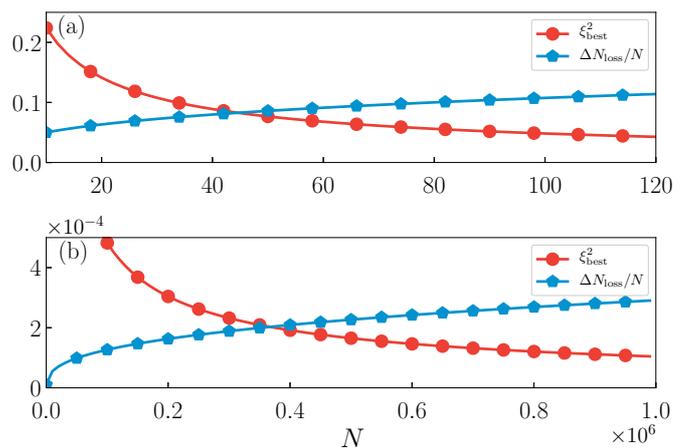}
\caption{
Best squeezing $\xi^{2}_{\rm best}=\xi^2(t_{\rm best})$ in the absence of decoherence and lost fraction due to 2-body and 3-body losses at the time $t_{\rm best}$ as function of the atom number in ${}^{87}$Rb for the transitions (a) $|1,1\rangle \leftrightarrow |2,-1\rangle$ with Feshbach tuned $a_{ab}=95a_0$ and dominant two body losses $K_{b}^{(2)} = 8.1 \times 10^{-20} m^3/s$ \cite{PhysRevA.84.021605} and $K_{ab}^{(2)} = 1.708 \times 10^{-19} m^3/s$ \cite{Opanchuk2012} and (b) $|1,-1\rangle \leftrightarrow |2,-2\rangle$ with three body losses only $K_{a}^{(3)} = 5.4 \times 10^{-42} m^6/s$ \cite{PhysRevLett.99.190402} and $K_{b}^{(3)} = 1.8 \times 10^{-41} m^6/s$ \cite{Soding1999}.}
\label{fig:losses}
\end{figure}

While two and three-body losses are suppressed in the Mott-insulator phase, they cannot be neglected in the superfluid phase, and play an important role as soon as the lost fraction of atoms at the best squeezing time becomes comparable to the squeezing $\xi_{\rm best}^2$ that one would have in the absence of decoherence \cite{PhysRevLett.100.210401}. For two particular configurations in Fig.~\ref{fig:losses} we show the lost fraction and the expected squeezing $\xi_{\rm best}^2$ in the absence of decoherence as a function of the initial atom number.
As the lost fraction increases with $N$, while $\xi_{\rm best}^2$  decreases,  the crossing of the two curves gives the maximum atom number $N_{\rm max}$ for which the decoherence due to atom losses can be neglected. The lost fraction is obtained by solving rate equations for the mean atom number
  \begin{align}
   \frac{d}{dt} \langle \hat{N}_\sigma \rangle(t) =  -\gamma_{\sigma}^{(2)}(t)  \, g_{\sigma}^{(2)}(t) \, \langle \hat{N}_\sigma \rangle^2(t) \nonumber \\
   - \gamma_{ab}^{(2)}(t) \, g_{ab}^{(2)}(t) \, \langle \hat{N}_a \rangle(t) \langle \hat{N}_b \rangle(t)
  \end{align}
 for two-body losses, and 
  \begin{equation}
   \frac{d}{dt} \langle \hat{N}_\sigma \rangle = - \gamma_{\sigma}^{(3)}(t) \, g_{\sigma}^{(3)}(t) \, \langle \hat{N}_\sigma \rangle^3
  \end{equation}
for three-body, where $\gamma_{\sigma}^{(m)}(t) = \frac{K_{\sigma}^{(m)} }{M^{m-1}}\int d^3 r\ w^{2m}(\mathbf{r},t)$ and $\gamma_{ab}^{(2)}(t) = \frac{K_{ab}^{(2)}}{2M} \int d^3 r\ w^4(\mathbf{r},t)$, with $K_{\sigma}^{(m)}$ the $m$-body loss rate constants for component $\sigma$,
$w(r)$ is the Wannier function identical for the $M$ sites, and $g_{\sigma}^{(m)}$ is
the on-site normalized $m$-body correlation function in our time-dependent lattice, for example 
$g_{a}^{(2)}=\langle a_i^\dagger a_i^\dagger a_i a_i \rangle/\langle a_i^\dagger a_i \rangle^2$,
calculated numerically in the adiabatic approximation using the Gutzwiller method.
In the case of the transition $|F=1,m_F=1\rangle \leftrightarrow |F=2,m_F-1\rangle$ for ${}^{87}$Rb atoms in Fig.~\ref{fig:losses}(a), where the interspecies scattering length $a_{ab}$ can be tuned by Feshbach resonance \cite{nature08919}, decoherence due to two-body losses is never negligible above $N_{\rm max} \approx 40$ atoms, and limits the squeezing to $10$dB already for a hundred atoms. On the other hand, 
for the transition $|F=1,m_F-1\rangle \leftrightarrow |F=2,m_F=-2\rangle$ in Fig.~\ref{fig:losses}(b), where the interspecies interaction can be tuned by slightly shifting the optical lattices for the two components \cite{PhysRevLett.82.1975} and two body losses are absent \cite{PhysRevA.80.042704}, the limit that is now imposed by three body losses is much less constraining and it does not influence the results up to large atom numbers $N_{\rm max} \approx 4 \times 10^5$.

\revision{In the Mott phase with one atom per site of a two-component system, two types of excitations exist: excitations that lead to double occupations, that are in general gapped,
and ``soft" excitations within the subspace of one atom per site that can be described by an effective spin model. In our scheme we need to maintain adiabaticity with respect to the 
first kind of excitations only, that is we need to remain in the manifold of single occupation. In this manifold, it is not important for us to be in the ground state, because the spin-dependent interaction energy 
is strongly reduced, scaling as $J^2/U$ \cite{SvistunovCF,Altman2CBH,Powel,HofstetterPRB} }\footnote{\revision{In our case the three interaction parameters $U_{\sigma\sigma'}$ have the same order of magnitude, $0<U_{ab} \lesssim \ U_{aa},U_{bb}$, so that we will refer to a generic $U$ in the following discussion.}}, \revision{and consequently the squeezing dynamics is practically stopped. 
In particular, contrarily to what happens in the superfluid phase, the presence of a gap separating the low energy manifold with one atom per site from the states with double occupation, allows us to raise the lattice to the deep Mott regime $J \to 0$ in a finite time, typically of order $\tau\geq\hbar/U$ \cite{NatuPRL}. One can estimate the effect of the residual squeezing dynamics during this time 
by comparing $\chi_{\rm resid} \tau$, where $\chi_{\rm resid}$ scales as the spin-dependent interaction eneregy per particle divided by $N$, with the expression of the adimensional squeezing time $T(t_{\rm best})\simeq N^{-2/3}$ below equation (\ref{eq:tb_scaling}). As
$\chi_{\rm resid} \tau \simeq (J/U)^2 N^{-1} \ll  N^{-2/3}$ we find that the residual squeezing dynamics during this adiabatic time is completely negligible. The only problem may come from the critical region where the gap closes as $\Delta/J \propto \sqrt{(J/U)_c-(J/U)}$ in the thermodynamic limit, scaling as 
$(\Delta/J)_c \propto N^{-1/3}$ at the transition in a finite size system \cite{CapogrossoPRB}. At the critical point we then find the same scaling $N^{1/3}$ of the adiabatic time as in the superfluid case. However, even taking $\tau=\hbar N^{1/3}/J$,  we would have $\chi_{\rm resid} \tau \simeq (J/U) N^{-2/3} \ll  N^{-2/3}$ meaning that the effect of the residual squeezing dynamics would still be small, due to the small value 
of $J/U$ near the critical point $(J/U)_c \simeq 0.03$.  }

\section{Conclusions}
We study the formation of a spin-squeezed atomic crystal by bringing an interacting two-component Bose-Einstein condensate across the superfluid-to-Mott transition in a time-dependent optical lattice. The scheme could be directly used in lattice clocks using a microwave transition.

\begin{acknowledgments}
\revision{We wish to thank Yvan Castin and Boris Svistunov for useful discussions}. This work was supported by the Polish National Science Center Grants
DEC-2015/18/E/ST2/00760, by the CNRS PICS-7403, and partially by the PL-Grid Infrastructure. D.K. acknowledges support from the French Government (BGF).
\end{acknowledgments}

\bibliographystyle{eplbib}
\bibliography{revised2}

\begin{thebibliography}{10}
\expandafter\ifx\csname url\endcsname\relax\def\url#1{\texttt{#1}}\fi

\bibitem{PhysRevLett.82.4619}
\Name{Santarelli G., Laurent P., Lemonde P., Clairon A., Mann A.~G., Chang S.,
  Luiten A.~N. \and Salomon C.} \REVIEW{Phys. Rev. Lett.}{82}{1999}{4619}.

\bibitem{nature12941}
\Name{Bloom B.~J., Nicholson T.~L., Williams J.~R., Campbell S.~L., Bishof M.,
  Zhang X., Zhang W., Bromley S.~L. \and Ye J.} \REVIEW{Nature}{506}{2014}{71}.

\bibitem{PhysRevA.81.023402}
\Name{Akatsuka T., Takamoto M. \and Katori H.} \REVIEW{Phys. Rev.
  A}{81}{2010}{023402}.

\bibitem{nphys1108}
\Name{Akatsuka T., Takamoto M. \and Katori H.} \REVIEW{Nat.
  Phys.}{4}{2008}{954}.

\bibitem{1882-0786-10-7-072801}
\Name{Takano T., Mizushima R. \and Katori H.} \REVIEW{Applied Physics
  Express}{10}{2017}{072801}.

\bibitem{PhysRevLett.101.220801}
\Name{Flambaum V.~V., Dzuba V.~A. \and Derevianko A.} \REVIEW{Phys. Rev.
  Lett.}{101}{2008}{220801}.

\bibitem{PhysRevLett.106.063002}
\Name{Chicireanu R., Nelson K.~D., Olmschenk S., Lundblad N., Derevianko A.
  \and Porto J.~V.} \REVIEW{Phys. Rev. Lett.}{106}{2011}{063002}.

\bibitem{Ludlow1805}
\Name{Ludlow A.~D., Zelevinsky T., Campbell G.~K., Blatt S., Boyd M.~M.,
  de~Miranda M. H.~G., Martin M.~J., Thomsen J.~W., Foreman S.~M., Ye J.,
  Fortier T.~M., Stalnaker J.~E., Diddams S.~A., Le~Coq Y., Barber Z.~W., Poli
  N., Lemke N.~D., Beck K.~M. \and Oates C.~W.}
  \REVIEW{Science}{319}{2008}{1805}.

\bibitem{Hinkley1215}
\Name{Hinkley N., Sherman J.~A., Phillips N.~B., Schioppo M., Lemke N.~D.,
  Beloy K., Pizzocaro M., Oates C.~W. \and Ludlow A.~D.}
  \REVIEW{Science}{341}{2013}{1215}.

\bibitem{JYe2017}
\Name{Campbell S.~L., Hutson R.~B., Marti G.~E., Goban A., Darkwah~Oppong N.,
  McNally R.~L., Sonderhouse L., Robinson J.~M., Zhang W., Bloom B.~J. \and Ye
  J.} \REVIEW{Science}{358}{2017}{90}.

\bibitem{PhysRevA.50.67}
\Name{Wineland D.~J., Bollinger J.~J., Itano W.~M. \and Heinzen D.~J.}
  \REVIEW{Phys. Rev. A}{50}{1994}{67}.

\bibitem{nature08988}
\Name{Riedel M.~F., B\"ohi P., Li Y., H\"ansch T.~W., Sinatra A. \and Treutlein
  P.} \REVIEW{Nature}{464}{2010}{1170}.

\bibitem{nature08919}
\Name{Gross C., Zibold T., Nicklas E., Est{\`e}ve J. \and Oberthaler M.~K.}
  \REVIEW{Nature}{464}{2010}{1165}.

\bibitem{PhysRevLett.104.073602}
\Name{Leroux I.~D., Schleier-Smith M.~H. \and Vuleti\'{c} V.} \REVIEW{Phys.
  Rev. Lett.}{104}{2010}{073602}.

\bibitem{Hosten2016}
\Name{Hosten O., Engelsen N.~J., Krishnakumar R. \and Kasevich M.~A.}
  \REVIEW{Nature}{529}{2016}{505} letter.

\bibitem{PhysRevLett.116.093602}
\Name{Cox K.~C., Greve G.~P., Weiner J.~M. \and Thompson J.~K.} \REVIEW{Phys.
  Rev. Lett.}{116}{2016}{093602}.

\bibitem{Greiner2002}
\Name{Greiner M., Mandel O., Esslinger T., Hansch T.~W. \and Bloch I.}
  \REVIEW{Nature}{415}{2002}{39}.

\bibitem{PhysRevA.79.041604}
\Name{Fukuhara T., Sugawa S., Sugimoto M., Taie S. \and Takahashi Y.}
  \REVIEW{Phys. Rev. A}{79}{2009}{041604}.

\bibitem{PhysRevA.77.011603}
\Name{Catani J., De~Sarlo L., Barontini G., Minardi F. \and Inguscio M.}
  \REVIEW{Phys. Rev. A}{77}{2008}{011603}.

\bibitem{ChinLett2009}
\Name{Zhou X.-J., Chen X.-Z., Chen J.-B., Wang Y.-Q. \and Li J.-M.}
  \REVIEW{Chin. Phys. Lett.}{26}{2009}{090601}.

\bibitem{PhysRevA.81.031611}
\Name{Lundblad N., Schlosser M. \and Porto J.~V.} \REVIEW{Phys. Rev.
  A}{81}{2010}{031611}.

\bibitem{PhysRevLett.117.150801}
\Name{Carr A.~W. \and Saffman M.} \REVIEW{Phys. Rev. Lett.}{117}{2016}{150801}.

\bibitem{PhysRevLett.112.150501}
\Name{Killoran N., Cramer M. \and Plenio M.~B.} \REVIEW{Phys. Rev.
  Lett.}{112}{2014}{150501}.

\bibitem{Bakr2009}
\Name{Bakr W.~S., Gillen J.~I., Peng A., Folling S. \and Greiner M.}
  \REVIEW{Nature}{462}{2009}{74}.

\bibitem{Sherson2010}
\Name{Sherson J.~F., Weitenberg C., Endres M., Cheneau M., Bloch I. \and Kuhr
  S.} \REVIEW{Nature}{467}{2010}{68}.

\bibitem{PhysRevA.67.052315}
\Name{van Loock P. \and Furusawa A.} \REVIEW{Phys. Rev. A}{67}{2003}{052315}.

\bibitem{PhysRevA.47.5138}
\Name{Kitagawa M. \and Ueda M.} \REVIEW{Phys. Rev. A}{47}{1993}{5138}.

\bibitem{Sorensen2001}
\Name{Sorensen A., Duan L.-M., Cirac J.~I. \and Zoller P.}
  \REVIEW{Nature}{409}{2001}{63}.

\bibitem{PhysRevLett.82.1975}
\Name{Jaksch D., Briegel H.-J., Cirac J.~I., Gardiner C.~W. \and Zoller P.}
  \REVIEW{Phys. Rev. Lett.}{82}{1999}{1975}.

\bibitem{PhysRevLett.81.5718}
\Name{Timmermans E.} \REVIEW{Phys. Rev. Lett.}{81}{1998}{5718}.

\bibitem{PhysRevLett.110.200406}
\Name{Gaunt A.~L., Schmidutz T.~F., Gotlibovych I., Smith R.~P. \and Hadzibabic
  Z.} \REVIEW{Phys. Rev. Lett.}{110}{2013}{200406}.

\bibitem{Zwirlein}
\Name{Mukherjee B., Z. Y., Patel P.~B., Hadzibabic Z., Yefsah T., Struck J.
  \and Zwierlein M.~W.} \REVIEW{Phys. Rev. Lett.}{118}{2017}{123401}.

\bibitem{Lundh2008}
\Name{Lundh E.} \REVIEW{The European Physical Journal D}{46}{2008}{517}.

\bibitem{PhysRevA.81.043620}
\Name{Wernsdorfer J., Snoek M. \and Hofstetter W.} \REVIEW{Phys. Rev.
  A}{81}{2010}{043620}.

\bibitem{PhysRevA.46.R6797}
\Name{Wineland D.~J., Bollinger J.~J., Itano W.~M., Moore F.~L. \and Heinzen
  D.~J.} \REVIEW{Phys. Rev. A}{46}{1992}{R6797}.

\bibitem{PhysRevA.95.063609}
\Name{Paw\l{}owski K., Fadel M., Treutlein P., Castin Y. \and Sinatra A.}
  \REVIEW{Phys. Rev. A}{95}{2017}{063609}.

\bibitem{PhysRevLett.107.060404}
\Name{Sinatra A., Witkowska E., Dornstetter J.-C., Li Y. \and Castin Y.}
  \REVIEW{Phys. Rev. Lett.}{107}{2011}{060404}.

\bibitem{Sinatra2012}
\Name{Sinatra A., Witkowska E. \and Castin Y.} \REVIEW{The European Physical
  Journal Special Topics}{203}{2012}{87}.

\bibitem{SvistunovCF}
\Name{Kuklov A.~B. \and Svistunov B.~V.} \REVIEW{Phys. Rev.
  Lett.}{90}{2003}{100401}.

\bibitem{Altman2CBH}
\Name{Altman E., Hofstetter W., Demler E. \and Lukin M.~D.} \REVIEW{New Journal
  of Physics}{5}{2003}{113}.

\bibitem{PhysRevB.44.10328}
\Name{Rokhsar D.~S. \and Kotliar B.~G.} \REVIEW{Phys. Rev. B}{44}{1991}{10328}.

\bibitem{PhysRevA.71.043601}
\Name{Zakrzewski J.} \REVIEW{Phys. Rev. A}{71}{2005}{043601}.

\bibitem{PhysRevLett.89.040402}
\Name{Jaksch D., Venturi V., Cirac J.~I., Williams C.~J. \and Zoller P.}
  \REVIEW{Phys. Rev. Lett.}{89}{2002}{040402}.

\bibitem{PhysRevE.78.066704}
\Name{Lim F.~Y. \and Bao W.} \REVIEW{Phys. Rev. E}{78}{2008}{066704}.

\bibitem{Sinatra1998}
\Name{Sinatra A. \and Castin Y.} \REVIEW{Eur. Phys. J. D}{4}{1998}{247}.

\bibitem{PhysRevA.57.3008}
\Name{Castin Y. \and Dum R.} \REVIEW{Phys. Rev. A}{57}{1998}{3008}.

\bibitem{PhysRevA.56.1414}
\Name{Gardiner C.~W.} \REVIEW{Phys. Rev. A}{56}{1997}{1414}.

\bibitem{PhysRevA.63.053601}
\Name{van Oosten D., van~der Straten P. \and Stoof H. T.~C.} \REVIEW{Phys. Rev.
  A}{63}{2001}{053601}.

\bibitem{PhysRevA.84.053613}
\Name{Zaleski T.~A. \and Kope\ifmmode~\acute{c}\else \'{c}\fi{} T.~K.}
  \REVIEW{Phys. Rev. A}{84}{2011}{053613}.

\bibitem{1751-8121-41-14-145005}
\Name{Ole\'s B. \and Sacha K.} \REVIEW{Journal of Physics A: Mathematical and
  Theoretical}{41}{2008}{145005}.

\bibitem{PhysRevLett.100.210401}
\Name{Li Y., Castin Y. \and Sinatra A.} \REVIEW{Phys. Rev.
  Lett.}{100}{2008}{210401}.

\bibitem{PhysRevA.84.021605}
\Name{Egorov M., Anderson R.~P., Ivannikov V., Opanchuk B., Drummond P., Hall
  B.~V. \and Sidorov A.~I.} \REVIEW{Phys. Rev. A}{84}{2011}{021605}.

\bibitem{Opanchuk2012}
\Name{Opanchuk B., Egorov M., Hoffmann S., Sidorov A.~I. \and Drummond P.~D.}
  \REVIEW{EPL}{97}{2012}{}.

\bibitem{PhysRevLett.99.190402}
\Name{Mertes K.~M., Merrill J.~W., Carretero-Gonz\'alez R., Frantzeskakis
  D.~J., Kevrekidis P.~G. \and Hall D.~S.} \REVIEW{Phys. Rev.
  Lett.}{99}{2007}{190402}.

\bibitem{Soding1999}
\Name{S\"{o}ding J., Guéry-Odelin D., Desbiolles P., Chevy F., Inamori H. \and
  Dalibard J.} \REVIEW{Appl. Phys. B}{69}{1999}{}.

\bibitem{PhysRevA.80.042704}
\Name{Tojo S., Hayashi T., Tanabe T., Hirano T., Kawaguchi Y., Saito H. \and
  Ueda M.} \REVIEW{Phys. Rev. A}{80}{2009}{042704}.

\bibitem{Powel}
\Name{Powell S.} \REVIEW{Phys. Rev. A}{79}{2009}{053614}.

\bibitem{HofstetterPRB}
\Name{Hubener A., Snoek M. \and Hofstetter W.} \REVIEW{Phys. Rev.
  B}{80}{2009}{245109}.

\bibitem{NatuPRL}
\Name{Natu S.~S., Hazzard K. R.~A. \and Mueller E.~J.} \REVIEW{Phys. Rev.
  Lett.}{106}{2011}{125301}.

\bibitem{CapogrossoPRB}
\Name{Capogrosso-Sansone B., Prokof'ev N.~V. \and Svistunov B.~V.}
  \REVIEW{Phys. Rev. B}{75}{2007}{134302}.

\end{thebibliography}

\end{document}